\documentstyle[12pt,epsfig,amssymb,citesort]{article}
\newcommand{\beq}{\begin{eqnarray}}
\newcommand{\eeq}{\end{eqnarray}}

\newcommand{\np}{Nucl. Phys.\ }
\newcommand{\pl}{Phys. Lett.\ }
\newcommand{\pr}{Phys. Rev.\ }
\newcommand{\asgen}{\alpha_s}

\newcommand{\as}{\alpha_{{}^{\widetilde{\rm MOM}}}}

\newcommand{\MSB}{\overline{\rm MS}}

\newcommand{\Lams}{\Lambda_{\overline{\rm MS}}}

\newcommand{\Gev}{{\rm GeV}}

\newcommand{\be}{\begin{equation}}
\newcommand{\ee}{\end{equation}}
\newcommand{\lwrsim}{\raise0.3ex\hbox{$<$\kern-0.75em\raise-1.1ex\hbox{$\sim$}}}

\def\np#1#2#3{Nucl.\ Phys.\ B#1 (19#3) #2}
\def\pl#1#2#3{Phys.\ Lett.\ #1B (19#3) #2}
\def\pr#1#2#3{Phys.\ Rev.\ D #1 (19#3) #2}

\def\prl#1#2#3{Phys.\ Rev.\ Lett.\ #1 (19#3) #2}

\begin{document}
\setcounter{page}{1}
\begin{flushright}
LPT-ORSAY 00/17\\
LTH 476\\
UPRF2000-05\\
RR003.0200\\
\end{flushright}
\begin{center}
\bf{\huge 
Lattice calculation of $1/p^2$ corrections
 to $\alpha_s$ and of $\Lambda_{\rm {QCD}}$  in
  the $\widetilde{MOM}$ scheme.}
\end{center}  
\vskip 0.8cm
\begin{center}{\bf  Ph. Boucaud$^a$, G. Burgio$^b$, F. Di Renzo$^{b,c}$, J.P. Leroy$^a$, 
\\
J. Micheli$^a$, C. Parrinello$^b$, O. P\`ene$^a$, C. Pittori$^d$, \\
J. Rodr\'\i guez--Quintero$^{a,e}$,  
 C. Roiesnel$^f$ and  K. Sharkey$^{c}$
}\\

\vskip 0.5cm 
$^{a}$ {\sl Laboratoire de Physique Th\'eorique~\footnote{Unit\'e Mixte 
de Recherche du CNRS - UMR 8627}\\
Universit\'e de Paris XI, B\^atiment 211, 91405 Orsay Cedex,
France}\\
$^b${\sl Dipartimento di Fisica, Universit\`a di Parma \\
     and INFN, Gruppo Collegato di Parma, Parma, Italy.}\\
$^c$ {\sl  Dept. of Mathematical Sciences, University of Liverpool\\
        Liverpool L69 3BX, U.K.}\\
$^d${\sl Dipartimento di Fisica, Universit\`a di Roma ``Tor Vergata'',
 Roma, Italy} \\
$^e${\sl Dpto. de F\'{\i}sica Aplicada e Ingenier\'{\i}a el\'ectrica \\
E.P.S. La R\'abida, Universidad de Huelva, 21819 Palos de la fra., Spain} \\
$^f$ {\sl Centre de Physique Th\'eorique\footnote{
Unit\'e Mixte de Recherche C7644 du CNRS} 
de l'\'Ecole Polytechnique\\
91128 Palaiseau Cedex, France }\\
\end{center}
\begin{abstract}

We report on very strong evidence of the occurrence of power terms 
in $\as(p)$, the  QCD running coupling constant in the $\widetilde{MOM}$ 
scheme, by analyzing non-perturbative measurements 
from the lattice three-gluon vertex between 2.0 and 10.0 GeV at zero flavor. While
 putting forward the caveat 
that this definition of the coupling is a gauge dependent one, 
the general relevance of such an occurrence is discussed. We fit 
$\Lambda_{\overline{\rm MS}}^{(n_f=0)}= 237 \pm 3\ ^{+\ 0}_{-10}$ MeV in perfect agreement
 with the result obtained by the
 ALPHA group with a totally different method.
  The power correction to  $\as(p)$ is fitted to 
 $(0.63\pm 0.03\ ^{+\ 0.0}_{-\ 0.13})\,{\rm GeV}^2/p^2$.

\end{abstract}

\section{Introduction}
\label{sec:intro}
 
The non-perturbative computation of the running QCD coupling constant
$\asgen(p)$ follows a two-sided goal: the large energy matching to perturbative
asymptotic QCD  formula turns out to be a most direct method to predict $\Lams$
from QCD first principles \cite{cpcp}. On the other hand, the moderate or low
energy behavior of $\asgen(p)$ is extremely instructive about non-perturbative
properties of QCD. In this paper we restrict ourselves to high and intermediate
energies and consider power corrections ($\sim 1/p^2$) to the leading 
asymptotic behavior. As we shall see it turns out that there is no sharp 
separation between the asymptotic domain and the intermediate one. The power
correction, beyond the lessons it contains by itself, greatly improves  our
asymptotic study and is never negligible up to $\sim 10$ GeV ! This surprising
fact could only be revealed thanks to the high accuracy achieved in the present
work.

The asymptotic approach has been recently followed in Ref. \cite{frenchalpha}.
It is worth   remarking that this matching procedure has also been developed for
the two-point Green function to  the same goal \cite{propag}, both matching
programs leading to a consistent estimate of $\Lams$. The almost two-sigma
discrepancy between the last estimate and the one obtained from Schr\"odinger
functional methods \cite{Lusch} seems to imply that some source of systematic
uncertainty remains uncontrolled. Furthermore the careful study of the
asymptotic behavior carried out for the gluon propagator in Refs.
\cite{propag} stresses a danger:  it is sometimes difficult to distinguish
between the {\it real asymptotic  scaling} and a behavior {\it mimicking
asymptoticity} but with a certain effective ``re-scaled''  $\Lambda$
parameter which differs  significantly from the real asymptotic one.
Consequently, and in spite of the agreement between both above-mentioned
estimates of $\Lams$  (matching two or three-point Green functions to
perturbative formulae), we must inquire whether both are not biased by some
sizable non-perturbative effect.

The operator product expansion (OPE) gives  a standard procedure to
pa\-ra\-me\-tri\-ze non-perturbative QCD effects  in terms of power corrections
to perturbative results. In this framework,  the powers involved in the
expansion are expected to be uniquely  fixed by the symmetries and the
dimensions of the operators  appearing in the product expansion. It should be
noted that,   due to the asymptotic nature of QCD perturbative  expansions,
power corrections are reshuffled between operators and  coefficient functions
in the OPE \cite{renormalons}.

Since we work in Landau gauge, the gauge dependent local operator $A_\mu A^\mu$
is allowed in the OPE \cite{lavelle} implying a dominant $1/p^2$ power
correction. Indeed sizable $1/p^2$ corrections are present as we shall show at
length in the next section. 

In a less straightforward way our result is related to another hot issue
(this is the spirit of the preliminary study in \cite{prevpow}): 
 the possible presence of $1/p^2$ terms in {\it gauge invariant} quantities
such as Wilson loops \cite{Ceccobeppe}.  Since no gluon local gauge invariant
operator of dimension less than 4 exists it is expected from OPE that the
dominant power correction should be $\propto 1/p^4$. 
This picture has however recently been challenged 
\cite{Akhoury,etc,Ceccobeppe}. It was pointed out that   power corrections
which are not {\it a priori} expected from the OPE  may in fact appear in the
expansion of physical observables.  Such terms may arise from (UV-subleading)
power corrections to  $\asgen(p)$, corresponding to non-analytical
contributions to the  $\beta$-function. It is worth stressing the following.
One  knows that the perturbative analysis does not encode all the information 
on the coupling. Among all that is missed, a {\it peculiar} contribution  to
the coupling could result in a {\it peculiar} correction to  physical
observables.  As a matter of fact, some evidence for an unexpected 
$\frac{\Lambda^2}{p^2}$ contribution to the gluon condensate  was obtained
through lattice calculations in Ref. \cite{Ceccobeppe}   (see also
\cite{Lepage} for an early evidence of such a contribution,  although the
perturbative series involved was not managed up to  high orders). 

Let us insist: there is no {\it direct} relation between the $1/p^2$
corrections found in the present work to $\asgen(p)$ in a {\it gauge dependent
scheme} and the power corrections advocated in the preceding paragraph  to
justify $1/p^2$ corrections to {\it gauge independent quantities}. Nobody
knows  how to relate a gauge dependent scheme to a gauge independent one beyond
the perturbative regime. Still, it may be that these two phenomena are not
completely unrelated and the large $1/p^2$ corrections found in
the present  work might trigger some further thoughts along this line.

The paper is organized as follows: in Section \protect\ref{sec:lattice}  we
explain the meaning of the lattice data, our strategy for the analysis and
report the results. In Section \protect\ref{sec:relev}  we briefly
review some theoretical arguments in support of power corrections  to $\asgen
(p)$, illustrating the special role of the 
$\frac{\Lambda^2}{p^2}$ term.  Finally, in Section \protect\ref{sec:conc} we draw
our conclusions. 

\section{Lattice calculation of $\alpha_s$ from Green functions
 and power corrections}
\protect\label{sec:lattice}

\subsection{$\asgen$ on the lattice} 

Several methods for computing $\alpha_s (p)$ non-perturbatively 
on the lattice have been proposed in recent years 
\cite{Aida,Nara,Bali,Michael,io}. In most cases, the goal of such 
studies is to obtain an accurate prediction for $\asgen (M_Z)$, 
i.e. the running coupling at the $Z$ peak, which is a fundamental 
parameter in the standard model. For this reason, lattice parameters 
are usually tuned so as to allow the computation of $\asgen (p)$ at 
momentum scales of at least a few \Gev s, where the two-loop asymptotic 
behavior is expected to dominate and power contributions are suppressed. 

As we shall see, however, non-perturbative power corrections cannot be 
neglected at energy scales as large as 10 GeV
which is a sufficient reason to consider them in the fit. As a bonus the
knowledge of these power corrections provides us with a physically significant
quantity as argued in the introduction. 

For this purpose, the best method is one where 
one can  measure $\asgen (p)$ in a wide range of momenta from a
single Monte Carlo data set. Indeed, a narrow energy window does not allow
to disentangle in a clear cut manner the power corrections from unknown
higher perturbative  orders, and these corrections can be mimicked by an
effective $\Lambda_{\rm{QCD}}$ different from the asymptotic one. 

One method which fulfills the above criterion is the determination of 
the coupling from the renormalized lattice three-gluon vertex 
function \protect\cite{io,cpcp,frenchalpha}.  
This is achieved by evaluating two- and three-point off-shell 
Green's functions of the gluon field in the Landau 
gauge, and imposing non-perturbative renormalization conditions on them, 
for different values of the external momenta. 
By varying the renormalization scale $p$, one can determine $\alpha_{s} (p)$ 
for different momenta from a single simulation. Obviously 
the renormalization scale must be chosen in a range of lattice momenta 
such that both finite volume effects and discretization  errors are under 
control. Both these constraints impose too strict limits on the energy range if
only one lattice run is used. 
This is why the procedure used in Ref. \cite{propag} combining several 
lattice simulations at 
different $\beta$'s is a necessity to get a larger range of lattice momenta. 
The use of different volumes will also help to control finite volume artifacts.
 
The definition of the coupling corresponds to a  
momentum-subtraction renormalization scheme in continuum QCD \cite{politzer}. 
It should be noted that in this scheme the coupling is a gauge-dependent 
quantity. As we already mentioned, one consequence of this fact is that 
$1/p^2$ corrections should be expected, based on OPE considerations.  
For full details of the method and the lattice calibration 
we refer the reader to Ref. \protect\cite{cpcp,frenchalpha}.

\subsection{Models for power corrections and construction of the data set} 
\protect\label{sec:modeldataset}

In the present work we shall not 
address the general problem  of defining an optimal analytic form for
 the coupling at all scales to which we could fit our data. 
For the purpose of our investigation, we shall compare the non-perturbative 
data for $\asgen$ with simple models obtained by adding a power correction 
term of the form $1/p^2$ to the perturbative formula at a given order. 
This amounts to a quite raw separation between {\it a perturbative versus a 
non--perturbative contribution}, the major problem of course being 
the possible interplay between a description in terms of (non-perturbative) 
power corrections and our ignorance about higher orders of perturbation 
theory. As crude as it may be, our recipe allows for addressing this problem, 
as we shall see. 

In order to identify a momentum interval where our ansatz could fit the data,  
one should keep in mind that the momentum range should start well above  the
location of the perturbative Landau pole, but it should nonetheless  include
low scales where power corrections may be large up to large enough  momentum
scales in order to  be confident that the asymptotic regime ({\it i.e.}
perturbative running)  has become dominant. Our choice will be\footnote{The
range's upper limit is determined by the condition $a \; p \le \pi/2$, where
$a$ is the lattice spacing, applied to our $24^4$ lattice at $\beta=6.8 \;$.}
2.0 GeV $\le p \le $ 9.6 GeV. It will turn out that in the {\it whole range}
both the perturbative three-loop contribution and the non perturbative $1/p^2$
one contribute by a sizable amount, although obviously the former becomes
dominant at larger scales.

A data set which fulfills the above-mentioned requirements can be constructed
by  aggregating data computed at different $\beta$ values ($\beta=$6.0, 
6.2, 6.4, 6.8) on a $24^4$ lattice. The fact that such
a data  set can be assembled is a very good {\it a posteriori} check that  the
expected scaling in the lattice cutoff $a$ takes place.

\begin{figure}
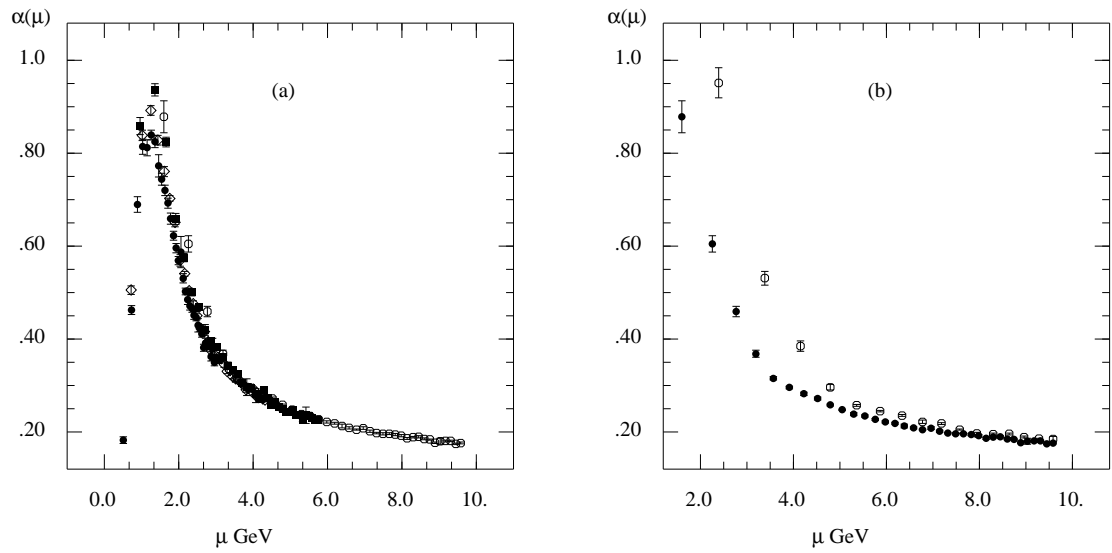

\hspace*{-1.3cm}
\begin{tabular}{ll}
\epsfxsize7.5cm\epsffile{Fig1a.eps} & 
\epsfxsize7.5cm\epsffile{Fig1b.eps} \\
\end{tabular}
\caption{\it {\small Evaluations of $\alpha_S$ from $24^4$ lattices at 
$\beta=6.0,6.2,6.4,6.8$ are respectively shown in plot (a) with black 
circles, white diamonds, black squares and white circles. In plot (b) 
black (white) circles correspond to $\alpha_S$ evaluations from a $24^4$ 
($16^4$) lattice at $\beta=6.8 \,$.}}
\label{Fig1}
\end{figure}

On the other hand, the physical volumes for these simulations are very different
from each other. The appropriate matching of the data proves that finite-size
effects remain controlled. The pattern for these volume effects is clear from
fig.~\ref{Fig1}(a), where the whole set of data is plotted, including the points
too much affected by finite volume artifacts and eventually rejected in our fits.
As can be seen, the effects are negligible for large enough  $L p$, where L is
the physical lattice length. Evidences for the same trend arise from the
comparison of data obtained on  two different lattices ($16^4$, $24^4$) at
$\beta=6.8$, shown in fig.~\ref{Fig1}(b).  In practice we will take as infrared
cut-off at each $\beta$ the value $p_{\min}$ such that, including points below
this value, the $\chi^2_{d.o.f}$ increases dramatically.  This criterion leads
to: $p_{min}(6.0)=2.0$ GeV, $p_{min}(6.2)=2.5$ GeV,  $p_{min}(6.4)=4.0$ GeV and 
$p_{min}(6.8)=6.0$ GeV. It corresponds to  $L p \gtrsim 24 $~\footnote{Almost all
the data from our $16^4$ lattice at $\beta=6.8$ turn out to be excluded by such a
requirement,  $L p \gtrsim 24$ leading for example to an infra-red cut-off equal
to 9. GeV. This is why we will not use at all the volume $16^4$.}.   Reversely, when we vary
the infrared cut-offs above the values just mentioned,  the $\chi^2_{d.o.f}$
stays  stable, in the range 1.4 to 1.6. Even more striking,  the resulting value for $\Lams$ is very stable, varying
no more than 2 MeV. Consequently the data set obtained  with these cut-offs
should be considered as IR safe and will be used  in the following fits.

At each $\beta$ value one should of course also worry of data in the 
range of the larger values of momentum, which are affected by lattice 
artifacts of $O(a^2p^2)$ (UV discretization effects). This was taken 
care of by the 
{\it sinus-improvement} program which has already been described in 
\cite{frenchalpha}, which basically amounts to the substitution of the 
lattice momenta $\frac{2\pi n_{\mu}}{L}$ with their $O(a^2p^2)$ analogues 
$\frac{2}{a}\sin(\frac{ap_{\mu}}{2})$. As already noticed in 
\cite{frenchalpha}, a good rationale for this in our gauge-fixed 
situation is that the gauge fixing algorithm leads to the relation 
$\frac{2}{a}\sin(\frac{ap_{\mu}}{2}) \, A_{\mu}(p) = 0$, while 
$p_{\mu} \, A_{\mu}$ does not vanish\footnote{One should keep in mind 
that we are imposing Landau gauge.}. It should be noticed that 
without this prescription the quality of almost any fit degenerates. 
 The relevant configurations were generated 
on a QH1 Quadrics system based in Orsay (see \cite{frenchalpha}).

\subsection{Fitting data to our ansatz} 
\label{fitting}

We now proceed to fitting the data set to our ansatz, which, we recall, is 
the addition of a term proportional to $\frac{1}{p^2}$ to a given order 
of perturbation theory:

\beq
\alpha_s(p^2)=\alpha_s^{Pert}(p^2)+ \frac{c}{p^2} \ .
\label{LaFor}
\eeq
 
\noindent Working at three loop level, the perturbative expression for the
running coupling  constant, $\alpha_S^{Pert}(p^2)$ requires to inverse either
the unexpanded formula 

\[
 \widetilde\Lambda=\widetilde\Lambda^{(c)}(\widetilde\alpha)\left(1+\frac 
 {\beta_1\widetilde\alpha}{2\pi\beta_0}+
 \frac
 {\widetilde\beta_2\widetilde\alpha^2}{32\pi^2\beta_0}\right)^{\frac{\beta_1}{2\beta_0^2}}
  \]
  \beq \times \exp
\left\{\frac{\beta_0\widetilde\beta_2-4\beta_1^2}{2\beta_0^2\sqrt{\Delta}}\left[
\arctan\left(\frac{\sqrt{\Delta}}{2\beta_1+\widetilde\beta_2\widetilde\alpha/4\pi}\right)
-\arctan\left(\frac{\sqrt{\Delta}}{2\beta_1}\right)\right]\right\}
 \label{lambda3}\eeq

\noindent or the expanded one

 \beq
 \widetilde\Lambda = \widetilde\Lambda^{(c)}(\widetilde\alpha)\left(1+\frac 
 {8\beta_1^2-\beta_0\widetilde\beta_2}{16\pi^2\beta_0^3}\widetilde\alpha
 \right)
  \label{lambdaexp}\eeq

\noindent where $\widetilde\Lambda^{(c)}$ denotes the conventional two loops formula:
 \beq       
             \widetilde \Lambda^{(c)} \equiv p \exp\left (\frac{-2 \pi}{\beta_0
	      \widetilde\alpha(p^2)}\right)\times
	      \left(\frac{\beta_0  \widetilde\alpha(p^2)}{4 \pi}\right)^{-\frac {\beta_1}
	      { \beta_0^2}}\label{lambda} \;\; ;
\label{LambdaConv}
\eeq

\noindent and $\Delta\equiv 2\beta_0\widetilde\beta_2-4\beta_1^2>0$ 
(for our $\widetilde{\rm{MOM}}$ 
scheme).
In the previous formula, the use of $\widetilde\Lambda$, $\widetilde\alpha$ 
and $\widetilde\beta$'s  stands for the $\Lambda$ parameter, the running 
coupling constant and beta function coefficients in the  particular 
$\widetilde{\rm{MOM}}$ renormalization scheme. From now on we will 
systematically convert $\widetilde \Lambda$ into $\Lams$ using 
\cite{cpcp,frenchalpha}
\beq
\Lams = \exp (-{70}/{66}) \,\widetilde\Lambda \simeq 0.346\,\widetilde\Lambda 
\label{raplambdas}
\eeq
In Eqs. (\ref{lambda3}-\ref{lambdaexp})  the $p^2$
dependence of $\widetilde\alpha$ has been omitted for simplicity.

No analytical expression can exactly inverse neither unexpanded 
eq.~(\ref{lambda3}) nor expanded eq.~(\ref{lambdaexp}). The following formula gives
an approximated solution to the inversion of eq.~(\ref{lambdaexp}): 

\begin{eqnarray}
\protect\label{3lp}
\widetilde{\alpha}(p^2) & = & \frac{4 \pi}{\beta_0 \, t} \, - \, 
\frac{8 \pi \beta_1}{\beta_0} 
\frac{\log(t)}{(\beta_0 \, t)^2} \nonumber \\
& &  + \, \frac{1}{(\beta_0 \, t)^3} \, 
\left( \frac{2 \pi \widetilde\beta_2}{\beta_0} + \frac{16 \pi \beta_1^2}{\beta_0^2} 
(\log^2(t) - \log(t) - 1 ) \right) 
\end{eqnarray}

\noindent where $t = \log(p^2/\widetilde{\Lambda}^2)$. 
On the other hand, an exact numerical
inversion of Eq.~(\ref{lambda3}) can be easily obtained and used to fit our data.
Both ans\"atze, eq.~(\ref{3lp}) and  the numerical inversion of eq.~(\ref{lambda3}),
should only differ by perturbative contributions higher than three loops. Thus, we
will fit with the two ans\"atze, the difference between these two  predictions being
considered as an estimate of the systematic uncertainty coming from  the neglected
perturbative orders. To make more direct the comparison with the Schr\"odinger 
functional estimate of $\Lams$ \cite{Lusch}, 
the central value for our prediction of $\Lams$
should be taken from  the fit with the exact inversion of Eq.
(\ref{lambda3})\footnote{The ALPHA collaboration estimate: $\Lams= 238 (19)$ MeV
 comes from
evaluating eq.~(\ref{lambda3}) for a value of $\alpha_S$ obtained from the 
lattice at very high momenta. }. This yields:

\beq\protect\label{res}
 \Lams  = 237 \pm 3 \ {\rm MeV},\quad
c=0.63 \pm 0.03 \ {\rm GeV}^2,\quad \chi^2/{\rm d.o.f.} = 1.6 \, \,  ; 
\eeq
in perfect agreement with the determination \cite{Lusch}
which uses a totally different technique.
Using eq.~(\protect\ref{3lp})
we obtain $\Lams = 227(5)\ {\rm MeV}$
and $c = 0.50(6)\  \rm{GeV}^2$. Comparing 
the latter results with eq.~(\protect\ref{res}) provides an estimate of the 
higher loop uncertainty of about 10 MeV for $\Lams$ and $0.1\  \rm{GeV}^2$  for $c$.  
 The size of  the power correction will be discussed in sections 3.1 and 3.2. 

\begin{figure}[hbt]
\begin{center}
\epsfxsize10cm\epsffile{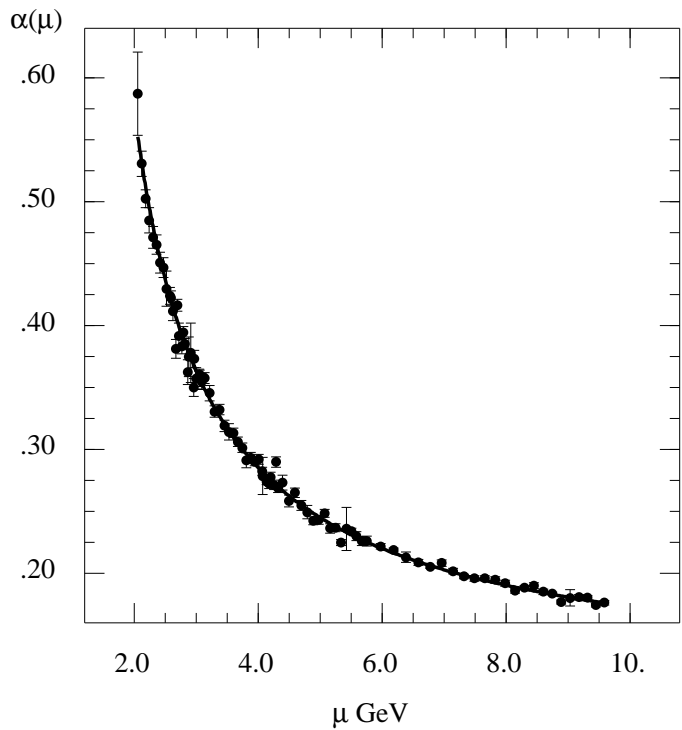} 
\caption{\it {\small Formula (\ref{LaFor}), with the values of $\Lams$ and $c$
given by eq.~(\ref{res}), fits impressively the data set built as 
explained in the text}}
\label{Fig2}
\end{center}
\end{figure} 

Fig. 3 illustrates the effect of the power correction 
 in a striking manner. The upper  set of
points shows $\Lambda$, converted to $\Lams$,
 computed through eq.~(\ref{lambda3}) 
from $\alpha_s$ provided by the lattice
at every value $p$. Scaling would imply a
constancy of $\Lams$ which is far
from true. The lower set of points corresponds to the same formula
applied to  $\alpha_s(p^2)^{\rm{lattice}}- 0.63/p^2$, i.e. to what should fit  the
perturbative formula. Now the constancy of $\Lams$ is very good within
the statistical errors. It is now clear why the upper points show  a trend to
decrease: as the energy scale increases, the effect of the power corrections must
decrease, and the upper points converge slowly towards the  lower ones. The
surprise is that above 9.0 GeV the merging of the two sets  of points has not yet
taken place contrarily to the general expectation that power corrections are
negligible at such a scale. We will elaborate on this in the next section.

We note that using eq. (\protect\ref{3lp}) one can draw the same conclusion: 
imposing $c=0$ (i.e. fitting to a pure three loop formula) there is no 
good fit on the whole range of momenta and the best that one can obtain 
on a restricted interval yields a value for $\Lams$ higher than expected 
(e.g. with respect to \cite{Lusch}) and a definitly worse $\chi^2_{d.o.f}$.


\begin{figure}[hbt]
\begin{center}
\epsfxsize8cm\epsffile{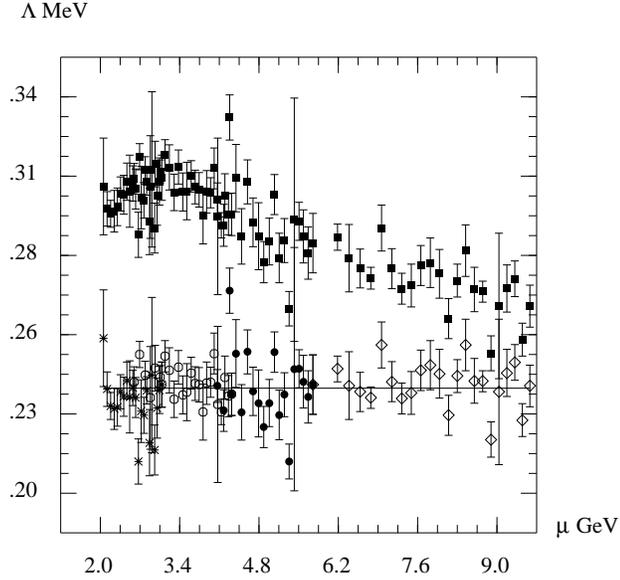} 
\caption{\small {\it The estimates of $\Lams$ computed from $\alpha_s$
via eq.~(\ref{lambda3}) are 
plotted as a function of the renormalization point momenta. Diamonds, black 
and white circles, and asterisks respectively correspond to evaluations from
$24^4$ lattices at $\beta=6.8,6.4,6.2,6.0$. Upper points are computed 
from the $\alpha_s$ values directly obtained from the lattice, while the lower points
use $\alpha_s^{Pert}$
as defined in
eq.~(\ref{LaFor}) with $c$ given by eq.~(\ref{res}).}}
\end{center}
\label{Fig3}
\end{figure} 

We now proceed to address the question of the stability of our previous 
result with respect to the inclusion of the fourth loop. 
To fit  the value of $\Lambda$, one would need
 $\beta_3$, the four loop coefficient in the perturbative expansion of the $\beta$-function. 
This coefficient is unknown but one can pin down a reasonable range of variation. 
Let us consider the ratio $b_3/b_2$ with
 $b_3=\beta_3/(4\pi)^3$ and $b_2=\beta_2/(2\pi)^2$. 
This ratio is larger than one in the only scheme $(\MSB)$ for which it is known.  
On the other hand we  expect that, for a reasonably small value of the coupling,
  the perturbative expansion of the $\beta$-function is still in the regime
  in which it seems to converge at four loops. 
  As a working hypothesis, we then suppose that the contribution to 
  the $\beta$-function coming from $\beta_3$ 
(that is, the one proportional to $\alpha^5$) is not too much larger than 
the one coming from $\beta_2$ at a typical value of $\alpha \sim 0.4$, 
implying $b_3/b_2 < 2.5$.
Actually we conservatively vary $b_3$ from 0 up to  $5 b_2$
in the following exploration of  a large range of values for $b_3$.

\begin{table}
\centering
\begin{tabular}{|c|c|c|c|c|c|c|c|c|c|c|}
\hline
\hline
& \multicolumn{4}{c|}{4 loops} & \multicolumn{6}{c|}{4 loops + power} \\
\cline{2-11}
$b_3/b_2$ & \multicolumn{2}{c|}{whole} & \multicolumn{2}{c|}{$>3$}& 
\multicolumn{3}{c|}{whole} & \multicolumn{3}{c|}{$>3$} \\ 
\cline{2-11}
& $\chi^2_{dof}$ & $\Lams$ & $\chi^2_{dof}$ & $\Lams$ 
& $\chi^2_{dof}$ & $\Lams$ & $c$ & $\chi^2_{dof}$ & $\Lams$ & $c$ \\
\hline
0 & 7.8 & 299 & 8.3 & 294 & 1.6 & 237 & 0.63 & 1.6 & 235 & 0.67 \\
1 & 7.1 & 284 & 6.0 & 287 & 1.8 & 238 & 0.56 & 1.6 & 235 & 0.61 \\
2 & 28  & 259 & 4.1 & 279 & 2.7 & 229 & 0.57 & 1.6 & 236 & 0.53 \\
3 & 104 & 227 & 3.0 & 270 & 4.7 & 215 & 0.65 & 1.6 & 238 & 0.45 \\
4 & 145 & 215 & 2.9 & 261 & 7.3 & 202 & 0.75 & 1.8 & 237 & 0.37 \\
5 & 157 & 209 & 4.4 & 252 & 10.2 & 190 & 0.84 & 2.4 & 231 & 0.37 \\
\hline
\hline
\end{tabular}
\caption{\small {\it A collection of fitted parameters obtained by 
imposing different values for the ratio $b_3/b_2$ defined in the text. 
``Whole'' refers to the whole energy window (2 GeV $< \mu < $10 GeV) 
and ``$>3$'' to a momentum range above $3$ GeV.}}
\label{Tab1}
\end{table}								 

First of all we try a pure four loop fit (that is without any power 
correction). We observe that there is no good fit on the whole range of momenta. 
If one tries to add again a term proportional to $\frac{1}{p^2}$ 
to the four loop perturbative expression, the following should be noted. 
Good fits are recovered either on the whole range of momenta for 
$b_3/b_2 \; \lwrsim \; 1$ or by discarding momenta below $3$ GeV as 
$b_3/b_2 \; \lwrsim \; 4$.  As far as the 
value of $\Lambda$ is concerned, it is really stable 
 when the fits are of good quality\footnote{Even for 
$b_3/b_2=5$, one could obtain a good quality fit ($\chi^2=1.8$) with 
$\Lams=238$, if momenta below $4$ GeV are now discarded.}. As for the 
coefficient of the non-perturbative 
term $\frac{1}{p^2}$, it is less stable. 
Results are summarized in Tab. \ref{Tab1}.
One can see 
 how the ``new player on the ground'', $\beta_3$, is 
strongly correlated to the coefficient of the power term. This coefficient is 
anyhow fully consistent with the value determined at three loop level, 
as long as  $b_3/b_2 \; \lwrsim \; 2$ 
{\it i.e.} when the asymptotic behavior is not too dramatically perturbed by the 
four-loop contribution.

Again, much the same holds when fitting to a formula for $\alpha_s$ as a 
function of momentum (like eq. (\protect\ref{3lp})) at the four loop level. 
In order to trust a four loop more than power corrections one would need 
both to discard lower momenta and to accept  excedingly large values for $b_3$. 

Of course, the perturbative knowledge of $\beta_3$ coefficient is 
unavoidable to get total confidence on our results {\it versus} higher 
loop orders inclusion. One should take this last analysis only as 
a preliminary check. Still it is another indication that things are 
working pretty well (that is, consistently) with respect to our theoretical 
prejudice. In any case, the value obtained for $\Lams$ is almost insensitive
to $\beta_3$.

We collect all the hints from these many counterproofs. 
To estimate the systematic errors we
use two methods: first we compare 
 the exact inversion of eq.~(\ref{lambda3}) and the use of eq.~(\ref{3lp})
  on our results, 
 second we use the results on the whole energy range 
 in table \ref{Tab1} with a reasonable $\chi^2$.
  We take the three loop result with 
formula (\ref{lambda3}) as our central  value
\beq\protect\label{resfin}
 \Lams = 237 \pm 3\  ^{+\ 0}_{-10} \ {\rm MeV},\quad
c=0.63 \pm 0.03\ ^{+\ 0.0}_{-\ 0.13} \ {\rm GeV}^2 
\eeq

The present analysis has improved over 
the one presented in \cite{cpcp,frenchalpha} by 
using a larger data set which provides  a  wider momentum window
and  taking  into account a power correction. 
If one tried to repeat the 
fits only in the range of the lower momenta as we did in \cite{cpcp,frenchalpha},
 there would be no clear cut 
indication for power corrections and the effective value for $\Lambda$ would 
turn out to be higher. What the upper momenta data really do is 
to single out the asymptotic value for $\Lambda$, while deviation from 
asymptotia in the lower momenta data asks for power correction rather 
than higher loops.

\section{Discussion on the $\frac{1}{p^2}$ corrections to $\asgen (p)$}
\protect\label{sec:relev}

Power corrections to $\asgen (p)$ can be shown to 
arise naturally in many physical schemes \cite{pino,maclep}.
The occurrence of such corrections 
cannot be excluded {\em a priori} in any renormalization scheme. 
Even more so, as previously stated, in a gauge dependent renormalization 
scheme as the $\widetilde {\rm MOM}$ discussed here. Clearly, 
the non-perturbative nature of such effects makes it very hard to 
assess their dependence on the renormalization scheme, which is 
only very weakly constrained by the general properties of the 
theory.

As discussed in the following, several arguments have been put forward 
in the past to suggest that a most likely candidate for a leading power 
correction to $\asgen(p)$ would be the same term of order ${\Lambda^2}/{p^2}$ 
we found. Furthermore, it is worth  emphasizing that 
this does not contradict the OPE expectation for a gauge dependent quantity.

\subsection{Static quark potential and confinement}
\label{sec:stat}
Consider the interaction of two heavy quarks in the static limit 
 (for a more detailed discussion see \cite{zakEQ}). 
In the one-gluon-exchange approximation, the static coulombic potential 
$V(r) $ can be written as
\begin{equation}
\protect\label{HQP}
V(r) \, \propto \, \alpha_s \ \int d^3k \,   
\frac{\exp^{i \vec{k} \cdot \vec{r}}}{|\vec{k}|^2}.
\end{equation}

Using standard arguments of renormalon analysis, one may consider a 
generalization of (\protect\ref{HQP}) obtained by replacing $\asgen$ 
with a running coupling:
\begin{equation}
\protect\label{HQP2}
V(r) \, \propto \, \int d^3k \, \alpha_s (|\vec{k}|^2) \,  
\frac{\exp^{i \vec{k} \cdot \vec{r}}}{|\vec{k}|^2}.
\end{equation}

The presence in $\alpha_s(k^2)$ of a power correction term of the form 
$ c/{k^2}$ would generate a linear confining piece $ K r$ in the 
 potential $V(r)$. It would of course be totally unjustified
 to take seriously our power correction in order to derive 
the linear potential. If we nevertheless perform this exercise to
have a feeling of the scales, we get  
 $K=2/3 c \simeq 0.4$ GeV$^2$, while the
   usual string tension is around $0.2$ GeV$^2$.
Note that a standard renormalon analysis of (\protect\ref{HQP2}) 
(see \cite{zakEQ} for the details) reveals contributions to the 
potential containing various powers of $r$, but a linear contribution 
is missing. This is a typical result of renormalon analysis:  
renormalons can miss important pieces of non-perturbative information.

\subsection{An estimate from another lattice method}

The lattice community has been 
made aware for some time of 
the possibility of non-perturbative contributions to the running 
coupling; for a clear discussion see \cite{ChrisLAT94}. 
Consider the ``force" definition of the running coupling:
\begin{equation}\label{force}
\alpha_{q\bar{q}}(Q) = \frac{3}{4} r^2 \frac{dV (r)}{dr} \;\;\;\;\;
(Q = \frac{1}{r}),
\end{equation}
where again $V(r)$ represents the static interquark potential.
By keeping into account the string tension contribution to $V(r)$, which 
can be measured in lattice simulations, one 
obtains a $1/Q^2$ contribution, whose order of magnitude is given 
by the string tension itself. Ironically, this term has been 
mainly considered as a sort of ambiguity, resulting in an 
indetermination in the value of $\alpha(Q)$ at a given scale. 
 From a different point of view, such a term could be interpreted as a 
clue for the existence of a $\frac{\Lambda^2}{p^2}$ contribution, and it 
also provides an estimate for the expected order of magnitude of it, at 
least in one (physically sound) scheme. The same na\"\i ve exercise than
in the preceding subsection leads from eq.~(\ref{force}) to 
$K = 4/3 c \simeq 0.8$ ${\rm GeV}^2 $. 

In order to make a deeper contact with what we are actually studying, 
one should of course keep in mind that all the
preceding arguments would be physically sounder in the Coulomb gauge
in which the notion of force has a clearer meaning. In Landau gauge
such a picture is far from clear. Furthermore, it is known that the 
interquark potential becomes close to linear at distances larger than 0.5 fm,
which corresponds to a momentum smaller than 0.4 GeV, while our $1/p^2$ fit
only apply above 2.0 GeV i.e. at distances smaller than 0.1 fm. 
Still, a couple of considerations are in order at this point. We note 
that the rough order of magnitude of the power term we found is the same as 
inferred from the simple arguments from the static potential (some 
$10^{-1} \ {\rm GeV}^2$), even if we are in a different (UV) regime. While this 
could turn out to be accidental, it is nevertheless intriguing to 
think of some sort of relation. In any case, one should nevertheless 
note that the power correction that we have obtained 
is rather large with respect to the common wisdom of non-perturbative 
effects being negligible at scales such as 10.0 GeV. 
A further study of the relation between the power 
correction and the confining potential is clearly needed.

\subsection{On the relation with the lattice gluon condensate puzzle.}

Having just stressed that the contribution of the power correction is 
rather large also at a scale such as $10.0 \;\; {\rm GeV}$, this is a good point to 
go back to the argument referring to the unexpected result of 
\cite{Ceccobeppe}. As already mentioned in the introduction, 
non-perturbative contribution to the running coupling can be advocated 
in order to explain the ${\Lambda^2}/{Q^2}$ contribution to the gluon 
condensate. The argument runs as follows (see \cite{Ceccobeppe}). From 
general arguments one expects the condensate $W$ to be written in the 
form

\beq \int_{0}^{Q^2} \frac{p^2 dp^2}{Q^4} 
\; f(\frac{p^2}{\Lambda^2})\; 
\eeq

\noindent which is based on the fact that this condensate has dimension 
four and is renormalization group invariant, so that the function 
$f({p^2}/{\Lambda^2})$, which is independent on $Q$ (for large $Q$), 
can be expressed as a function of a running coupling. This leads to 
consider the contribution coming from the large frequency part of 
the integral 

\begin{equation}
\protect\label{intalfa}
\int_{\rho \Lambda^2}^{Q^2} \frac{p^2 dp^2}{Q^4} 
\; \asgen(\frac{p^2}{\Lambda^2})\; 
\end{equation}

\noindent in which the function $f(p^2/{\Lambda^2})$ is taken 
proportional to the running coupling\footnote{Note that higher powers 
would be subleading both for renormalons and for power corrections.}. 
By taking into account the perturbative running coupling the IR renormalon 
contribution can be obtained (see \cite{Ceccobeppe} for details). Again, 
one can consider also contributions coming from a non--perturbative 
correction to the coupling of the form $c/{p^2}$. From the UV 
limit of integration one then obtains a ${\Lambda^2}/{Q^2}$ 
contribution to $W$. Let us insist: this is a contribution coming from a 
coupling with a $c/{p^2}$ correction in the UV region. While 
stressing again that one can draw no definite conclusion from the 
following consideration, still it is worth noting that our scheme provides 
an example of a coupling in which a $c/{p^2}$ correction is 
not negligible even at $10.0 \;\; {\rm GeV}$. For a discussion of possible scenarios 
for the result of \cite{Ceccobeppe} see also \cite{Beneke}.

\subsection{Landau pole and analyticity.}

It is well known that perturbative QCD formulae for the running of 
$\asgen$ inevitably contain singularities, which are often referred 
to as the Landau pole. The details of the analytical structure depend on 
the order at which the $\beta$-function is truncated and on the 
particular solution chosen. The existence of an interplay between the 
analytical structure of the perturbative solution and the structure of 
non-perturbative effects has been advocated for a long time 
\cite{RedBog}. To illustrate this idea, consider the 
one-loop formula for the running coupling  $\alpha_s (p)$:
\begin{equation}
\alpha_s(p^2)~=~
{1\over b_0 \ \log(\frac{p^2}{\Lambda^2})}.
\end{equation}
Here the singularity is a simple pole,  which can be removed if one 
redefines $\alpha_s (p)$ according to the following prescription:
\begin{equation}
\alpha_s(p^2)~=~
{1\over b_0 \ \log(\frac{p^2}{\Lambda^2})}+
{\Lambda^2 \over b_0 
(\Lambda^2-p^2)}, 
\end{equation}
where a power correction of the asymptotic form $\frac{\Lambda^2}{p^2}$ 
appears. However, the sign of such a correction is the opposite of 
what one would 
expect from 
the results of \cite{Ceccobeppe} and from the results in Section
\ref{sec:lattice} (although the absolute value is of 
the right order),
 so that in the end one could envisage a more general 
formula for the regularized coupling: 
\begin{equation}
\alpha_s(p^2)~=~
{1\over b_0 \ \log(\frac{p^2}{\Lambda^2})}+
{\Lambda^2 \over b_0 
(\Lambda^2-p^2)}+\eta \ {\Lambda^2 \over p^2}. 
\label{eq:regu}
\end{equation}
The message from (\ref{eq:regu}) is that 
 the coefficient of the power correction is not constrained
 by the mere cancellation of the pole.

At higher perturbative orders one encounters multiple singularities, 
which include an unphysical cut. There are several ways to regularize 
them. In particular, the method discussed in \cite{RedBog} combines a 
spectral-representation approach with the Renormalization Group. The 
method was originally formulated for QED, but it has recently been 
extended to the QCD case \cite{Shirk}.

\section{Conclusions}
\protect\label{sec:conc}

We have studied the strong coupling constant estimated
non perturbatively on the lattice from Green functions 
in the Landau gauge using the $\widetilde {\rm MOM}$ scheme.
This has been performed with a large statistics of 1000
field configurations per run and running at $\beta=
6.0, 6.2, 6.4, 6.8$. Finite volume effects as well
as finite lattice spacing effects have been carefully controlled. 

We have parametrized the momentum dependence of $\alpha_s$
using the three-loop perturbative formula plus a $c/p^2$ term.
We have obtain a good fit in the energy range from 2.0 GeV
up to 10.0 GeV.
As a result of this study we find 

\beq\protect\label{resfin2}
 \Lams \ = 237 \pm 3\ ^{+\ 0}_{-10}\ {\rm MeV},\quad
c=0.63 \pm 0.03 \ ^{+\ 0.0}_{-\ 0.13} \ {\rm GeV}^2,
\eeq 

$\Lams$ agrees perfectly well with the result of \cite{Lusch}
and the existence of sizable power corrections is convincingly 
established. The stability of our fit has been 
extensively checked.

The power correction turns out to be rather large, 
providing a 3\% correction on $\alpha_s$ at 10.0 GeV,
i.e. a 20 \% correction on $\Lambda$.

Having gathered evidences for a particular scheme, one needs to address 
the issue of the general relevance of such a finding in the spirit of the 
discussion of section \protect\ref{sec:relev}, assessing the 
scheme dependence of our results. As already discussed, the 
non-perturbative nature of power corrections makes it very hard to 
formulate any theoretical procedure to estimate the impact of scheme 
dependence. 
The best one can do at this stage is to consider 
different renormalization schemes and definitions of the coupling and 
gather numerical evidence and 
formal arguments supporting power corrections to $\asgen(p)$.
In this way, scheme-independent features may eventually be identified. 
For example, on the basis of our results, we note the following:
\begin{itemize}
\item Theoretical arguments suggest $1/p^2$ corrections both for 
the coupling as defined from the static potential and for the one 
obtained from the three-gluon vertex. The arguments for the former case 
were outlined in Sections 3.1 and 3.2. As far as the coupling from the 
three-gluon vertex is concerned, $1/p^2$ corrections appear in an 
OPE analysis if one keeps into account the fact that for such a gauge dependent
 coupling dimension 2 condensates are expected. 

\item In the static potential case, the theoretical arguments also 
provide an estimate for the order of magnitude of the coefficient 
of the $1/p^2$ correction  while in the three-gluon vertex case 
the OPE arguments do not, suggesting instead that it
may depend on the gauge.  
Nevertheless , the  order of magnitude of our numerical result in the Landau 
gauge is roughly the same as the one from the static potential 
case. This calls 
for further investigation, which may be performed for example by attempting 
a similar calculation in a different gauge and particularly in the 
Coulomb gauge in which the static potential is naturally defined. 
\end{itemize}

This issue of scheme dependence could be the focus of possible future work.

\section{Acknowledgements} 
We thank Alain Le Yaouanc and Chris Michael
for stimulating discussions. J. R-Q is indebted to Spanish 
Fundaci\'on Ram\'on Areces for his financial support.
F. DR acknowledges 
both support from PPARC and from MURST (contract 9702213582) and INFN 
({\it i.s.} PR11). C. Pi. warmly thanks the ``Groupe de Phys. Nucl. Th. de
l'Univ. de Li\`ege'' for kind hospitality and 
acknowledges the partial support of IISN.     
These calculations were performed on the QUADRICS QH1 located 
in the Centre de Ressources
 Informatiques (Paris-sud, Orsay) and purchased thanks to a
  funding from the Minist\`ere de
  l'Education Nationale and the CNRS.


\begin{thebibliography}{9}
\bibitem{cpcp}
B. All\'es, D. S. Henty, H. Panagopoulos, C. Parrinello, C. Pittori, D. G. 
Richards,  Nucl. Phys. { B502} (1997) 325.
\bibitem{frenchalpha}Ph. Boucaud, J. P. Leroy, J. Micheli, O. Pene, 
C. Roiesnel, JHEP 9810 (1998) 017; JHEP 9812 (1998) 004.
\bibitem{propag}D. Becirevic, Ph. Boucaud, J. P. Leroy, J. Micheli,
 O. Pene, J. Rodriguez-Quintero, C. Roiesnel, Phys. Rev. { D60} (1999)
  094509;  hep-ph/9910204 (to be published in Phys. Rev. D.)
\bibitem{Lusch}
	S. Capitani, M. Guagnelli, M. L\"uscher, S. Sint, R. Sommer, 
	P. Weisz and H. Wittig,
	Nucl. Phys. Proc. Suppl. { 63} (1998) 153;
	Nucl. Phys. { B544} (1999) 669.
\bibitem{renormalons}
         For reviews and classic references see:\\
         V.I. Zakharov, Nucl. Phys. { 385} {1992} {452};\\
         A.H. Mueller, in {\it QCD 20 years later}, vol.~1 
         (World Scientific, Singapore 1993).
         B. Lautrup, \pl{69}{109}{77};
         G. Parisi, \pl{76}{65}{77}; \np{150}{163}{79};
         G. t'Hooft, in {\it The Whys of Subnuclear Physics}, Erice
         1977, ed A. Zichichi, (Plenum, New York 1977);
         M. Beneke and V.I. Zakharov, \pl{312}{340}{93};
         M. Beneke \np{307}{154}{93};
         A. H. Mueller, \np{250}{327}{85}; \pl{308}{355}{93};
         G. Grunberg, \pl{304}{183}{93}; \pl{325}{441}{94}.
\bibitem{lavelle} For a discussion of this issue about gluon propagators,
see for example M. Lavelle and M. Oleszczuk, Mod. Phys. Lett. { A 7} (1991)
3617.
\bibitem{prevpow}
	G. Burgio, F. Di Renzo, C. Parrinello and C. Pittori
	Nucl. Phys. Proc. Suppl. 73 (1999) 623, 
	Nucl. Phys. Proc. Suppl. 74 (1999) 388. 
\bibitem{Ceccobeppe}
	G Burgio, F. Di Renzo, G. Marchesini and E. Onofri, 
	\pl{422}{219}{98}. 
\bibitem{Akhoury}
        R. Akhoury and V.I. Zakharov, hep-ph/9705318.
\bibitem{etc} G.Grunberg, hep-ph/9705290, Presented at Rencontres de Moriond
 97 on  QCD and High Energy Hadronic Interactions; 
  JHEP 11 (1998) 006.
\bibitem{Lepage}
	G.P. Lepage and P. Mackenzie,
	Nucl. Phys. Proc. Suppl. 20 (1991) 173. 
\bibitem{Aida}
A.X.\ El-Khadra et al., Phys.\ Rev.\ Lett.\ { 69} (1992) 729.
\bibitem{Nara}
M.\ L\"uscher et al., Nucl.\ Phys.\ { B413} (1994) 481.
\bibitem{Bali}
G.S.\ Bali and K.\ Schilling, Phys.\ Rev.\ { D47} (1993) 661.
\bibitem{Michael}
S.P.\ Booth et al., Phys.\ Lett.\ { B294} (1992) 385.
\bibitem{io}
C.\ Parrinello, Phys.\ Rev.\ { D50} (1994) 4247.
\bibitem{politzer} See for example H.D. Politzer, Phys. Reports
 { 14C} (1974) 141.
\bibitem{pino}
         Yu.L. Dokshitzer, G. Marchesini and B.R. Webber,
          Nucl.\ Phys. { 469} {(1996)} {96}
\bibitem{maclep} S.J. Brodsky, G.P. Lepage and P.B. Mackenzie,
\pr{28} {228} {83}.
\bibitem{zakEQ}
	R. Akhoury and V.I. Zakharov, Phys. lett. { B438} (1998) 165.
\bibitem{ChrisLAT94}
	C. Michael, Nucl. Phys. Proc. Suppl. 42 (1995) 147.
\bibitem{Beneke}
	M. Beneke, Phys. Rept. 317 (1999) 1;   hep-ph/0001134 
	and references therein. 
\bibitem{RedBog}
	P. Redmond, \pr{112}{1404}{58};
	N.N. Bogoliubov, A.A. Logunov and D.V. Shirkov,
	Sov. Phys. JETP 37 (1959) 805.
\bibitem{Shirk}
	D.V. Shirkov and I.L. Solovtsov, 
	\prl{79}{1209}{97}.
\end{thebibliography}
\end{document}